\documentclass[%
 reprint,
superscriptaddress,
 amsmath,amssymb,
 aps,
]{revtex4-2}

\usepackage{graphicx}
\usepackage{dcolumn}
\usepackage{bm}
\usepackage{hyperref}
\usepackage{xcolor}

\begin{document}

\preprint{APS/123-QED}

\title[Anomalous Radiation Emission from Self-Modulated Plasma Mirrors\\]{Anomalous Relativistic Emission from Self-Modulated Plasma Mirrors\\}

\author{M. Lamač*}
\affiliation{ 
ELI Beamlines Facility, The Extreme Light Infrastructure ERIC, Za Radnicí 835, Dolní Břežany, 25241, Czechia
}%
\affiliation{ 
Faculty of Mathematics and Physics, Charles University, Ke Karlovu 3, Prague 2, 12116, Czechia
}%

\author{K. Mima}

\affiliation{ 
Institute of Laser Engineering, Osaka University, 2-6 Yamadaoka, Suita, Osaka 565-0871, Japan
}%

\author{J. Nejdl}
\affiliation{ 
ELI Beamlines Facility, The Extreme Light Infrastructure ERIC, Za Radnicí 835, Dolní Břežany, 25241, Czechia
}%

\affiliation{ 
Faculty of Nuclear Science and Physical Engineering, Czech Technical University in Prague, Břehová 7, Prague 1, 11519, Czechia}

\author{U. Chaulagain}
\affiliation{ 
ELI Beamlines Facility, The Extreme Light Infrastructure ERIC, Za Radnicí 835, Dolní Břežany, 25241, Czechia
}%

\author{S. V. Bulanov}%
\affiliation{ 
ELI Beamlines Facility, The Extreme Light Infrastructure ERIC, Za Radnicí 835, Dolní Břežany, 25241, Czechia
}%

\affiliation{Kansai Photon Science Institute, National Institutes for Quantum and Radiological Science and Technology, 8-1-7 Umemidai, Kizugawa, 619-0215, Kyoto, Japan}

\date{\today}

\begin{abstract}

The interaction of intense laser pulses with plasma mirrors has demonstrated the ability to generate {\color{black}high-order harmonics,} producing a bright source of extreme ultraviolet (XUV) radiation and attosecond pulses. Here, we report an unexpected transition in this process. {\color{black} We show that the loss of spatio-temporal coherence in the reflected high-harmonics can lead to} a new regime of highly-efficient coherent XUV generation, with an extraordinary property {\color{black}where the radiation is directionally anomalous, propagating parallel to the mirror
surface.} With analytical calculations and numerical particle-in-cell simulations, we discover that the {\color{black} radiation emission is due to} laser-driven oscillations of relativistic electron {\color{black}nanobunches which originate from a plasma surface instability.}
\end{abstract}

\maketitle

\begin{figure}
\includegraphics[scale = 0.23]{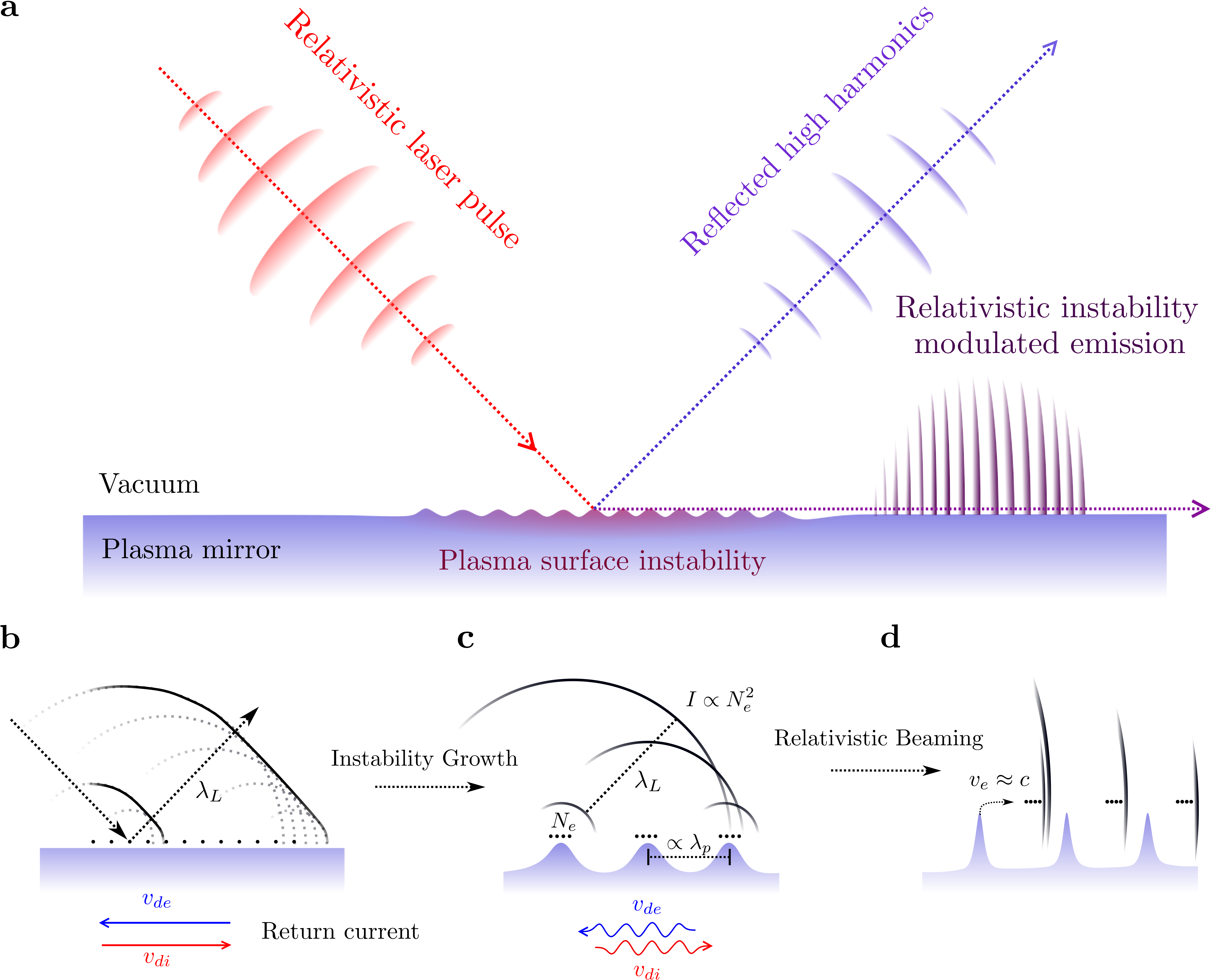}\caption{\textbf{Principle of RIME.}}
\label{fig:1}
\end{figure}

Generation of attosecond pulses and extreme ultraviolet radiation (XUV) originating from the interaction of a relativistically intense laser with a plasma mirror has now been investigated for almost three decades \citep{mori1993relativistic, teubner2009high, bulanov1994interaction, vincenti2014optical, pirozhkov2012soft}. With experimental results supporting its potential for applications in fundamental science, ultrafast science, attosecond interferometry or imaging \citep{teubner2009high, dromey2009diffraction, yeung2014dependence, heissler2012few, naumova2004relativistic, wheeler2012attosecond, smirnova2009high, neutze2000potential}, this mechanism presents a competitive alternative to the contemporary workhorse of attosecond science, which is high-harmonic generation (HHG) from noble gases originating from non-linear laser-atom interactions \citep{krausz2009attosecond, smirnova2009high, nefedova2017development, heyl2016scale}. The potential of relativistic high-harmonic generation reveals itself in the lack of ionization thresholds, which pose a hard limit on intensity of conventional gas or sollid HHG sources. With the growing availibility of commercial high-repetition rate laser systems with peak powers in the terawatt range, relativistic generation of broadband, bright and ultrashort radiation will be attractive even at small university-scale laboratories. The emission of relativistic high harmonics originates from plasma surface electrons {\color{black}performing non-linear oscillations} within the laser field, with velocity close to the speed of light. This occurs when the amplitude of the incident laser exceeds the threshold amplitude of relativistic optics \citep{mourou2006optics}, which is characterized in terms of the normalized laser amplitude as $a_{0} = eE_{0}/m_{e}\omega _{0}c \geq 1$, where $e$ is the elementary charge, $m_{e}$ the electron mass, $c$ the speed of light, $\omega_{0}$ is the angular frequency of the laser and $E_{0}$ the amplitude of the laser electric field. {\color{black}The coherent reflection from the plasma surface can be additionally affected for very high laser intensities} such that the ions can accelerate to a fraction of the speed of light in a single laser cycle, $a_{0} \geq (m_{i}/Zm_{e})^{1/2}$ \citep{macchi2005laser, esirkepov1999ion}, where $m_{i}$ is the ion mass and $Z$ is the ion charge number.

In this letter, we report the discovery of a new {\color{black} regime of XUV light generation}, which we call relativistic instability-modulated emission (RIME), originating from the interaction of an intense laser pulse with a plasma mirror, {\color{black}which is self-modulating due to unstable return current induced by the collisionless absorption of the laser}. RIME generates {\color{black}broadband XUV radiation with high-efficiency} and an anomalous propagation characteristic, where the radiation is emitted parallel to the plasma mirror surface. The mechanism of RIME is illustrated in Fig. \ref{fig:1}. An intense P-polarized laser pulse is obliquely incident on the surface of a relativistically overdense plasma mirror, {\color{black}$1 \leq a_{0} < n_{e}/n_{c}$}, where $n_{e}$ is the electron plasma density, $n_{c} = m_{e}\epsilon_{0}\omega_{0}^{2}/e^{2}$ is the critical plasma density and $\epsilon_{0}$ is the vacuum permittivity. {\color{black}At the beginning of interaction}, the intense incident laser {\color{black}coherently reflects from and} accelerates the {\color{black}surface} electrons into the plasma bulk (Fig. \ref{fig:1}b) by the Brunel effect \citep{brunel1987not, gibbon2005short}. These bulk-penetrating Brunel electrons induce neutrality-restoring return currents flowing along the mirror periphery, consisting of counter-streaming electrons and ions which generate intense quasi-static magnetic field on the mirror surface \citep{nakamura2004surface}. {\color{black}The electron-ion two-stream grows unstable} with a growth rate $\Gamma \propto \omega_{pe}(Zm_{e}/m_{i})^{1/3}$ \citep{buneman1959dissipation}, which is of the order of the laser frequency for solid high-Z targets with electron plasma densities $n_{e}/n_{c} \geq (m_{i}/Zm_{e})^{2/3}$. Therefore, {\color{black}after laser-plasma interaction time of $\approx 1/\Gamma$}, the plasma mirror self-modulates with wavelength of the order of the plasma wavelength $\lambda_{p} = 2\pi c/\omega_{pe}$, where $\omega_{pe} = \left( e^{2}n_{e}/m_{e}\epsilon_{0}\right)^{1/2}$ is the electron plasma frequency. {\color{black}As the plasma wave grows, the oscillating electrons are bunched and the individual bunch emissions are coherently-enhanced while the reflected wave loses its spatio-temporal coherence (Fig. \ref{fig:1}c). As the laser amplitude ramps up in the interaction region, the unstable plasma wave grows non-linear and breaks \citep{bulanov1978tearing}, releasing electron nanobunches to be accelerated by the laser across the magnetized plasma surface to velocities close to the speed of light, resulting in surface-parallel relativistic beaming of individual bunch emissions (Fig. \ref{fig:1}d)}.

The coherent intensity spectrum of RIME generated over a half-oscillation period by an electron bunch containing $N_{e}$ electrons can be obtained as (See Supplementary Material \citep{lamac2022supp} for derivation details) \begin{equation}
    I(\omega) = N_{e}^{2}\frac{\sqrt{3}e^{2}\gamma }{4\pi \epsilon_{0}c}\frac{\omega/\omega_{c}}{\left(1 + (\omega /\omega_{b})^{2}\right) ^{2}}  \int_{\omega/\omega_{c}}^{\infty}K_{5/3}(\xi )\hspace{0.5 mm}\text{d}\xi ,
\label{eq:1}
\end{equation}
where $K_{5/3}(x)$ is the modified Bessel function, $\omega_{c} = 3c\gamma^{3}/2\lambda_{0}$ is the critical frequency indicating an exponential cut-off in the spectrum, $\lambda_{0}$ is the laser wavelength and $\gamma $ is the electron Lorentz factor. The electron bunch modulation frequency is given as $\omega _{b} = c/L_{b}$, where $L_{b}$ is the electron bunch length. Bunch modulation frequency indicates a transition in the radiation spectrum, from the coherent regime of radiation emission, where $I \propto N_{e}^{2}$ for frequencies smaller than or of the same order $\omega \lesssim \omega_{b}$, into the regime of incoherent emission, where $I \propto N_{e}$ for $\omega \gg \omega_{b}$. 

\begin{figure}
\includegraphics[scale = 0.5]{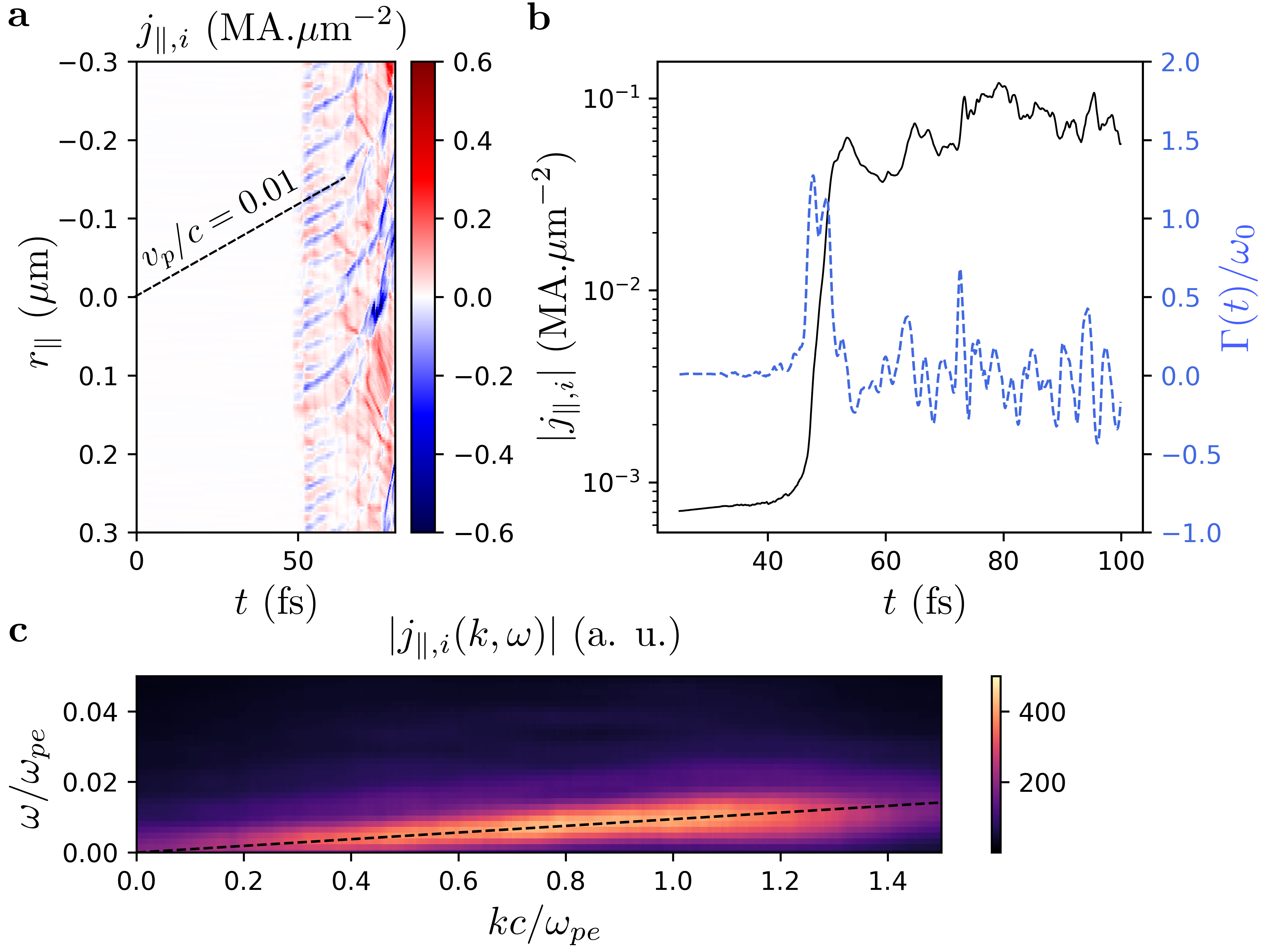}
\caption{{\color{black}\textbf{PIC simulation of the surface instability.} a) Time evolution of the longitudinal component of the ion current, shown up to $t = 80$ fs. Dashed line following linear phase velocity of the surface wave corresponds to $v_{p}/c = 0.01$. b) Time evolution of $|j_{i, \parallel}|$ averaged over $ r_{\parallel} \in [-1\ \mu\mathrm{m}, 1\ \mu\mathrm{m}]$ (black) and corresponding instantaneous growth rate $\Gamma (t)$ (blue). c) Fourier transform of the ion current in the linear phase of the instability, $0 \text{ fs} < t < 55 \text{ fs}$ {\color{black}(coloured, smoothened)} and the real part of the solution of Eq. \ref{eq:2} (black, dashed) calculated for the electron drift velocity $v_{de}/c = 0.01,\ \omega_{p}/\omega_{0} = 31.6$.}} 
\label{fig:2}
\end{figure}

\begin{figure*}
\includegraphics[scale = 0.32]{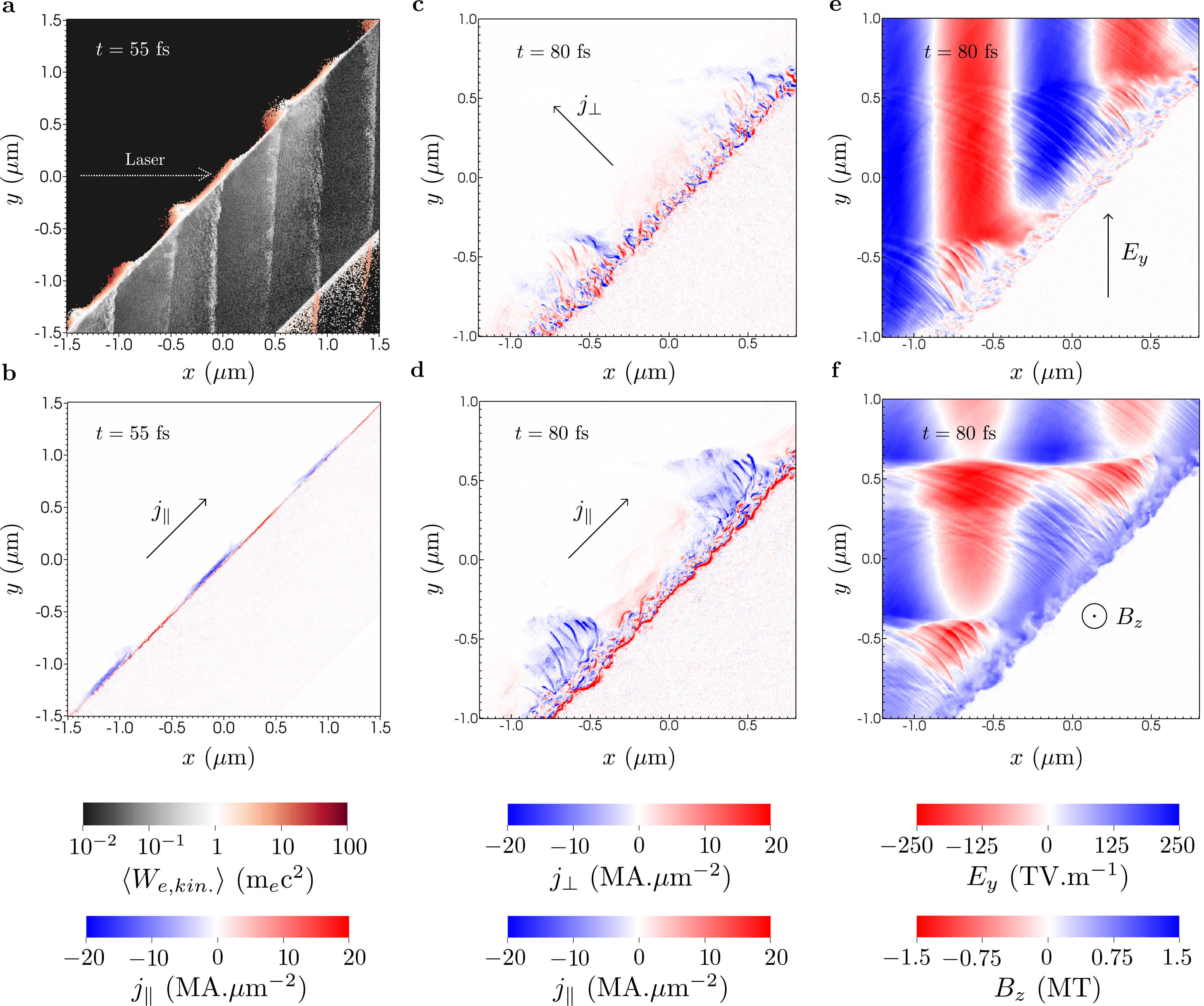}
\caption{{\color{black}\textbf{PIC simulation of RIME.} Average kinetic energy of electrons (a) and the longitudinal electron current (b) at the time of instability saturation, $t = 55$ fs. Transverse (c) and longitudinal (d) electron current, and electric (e) and magnetic field (f) components at the time of peak laser amplitude arrival, $t = 80$ fs.}}
\label{fig:3}
\end{figure*}

To describe the modulation frequency of the electron bunch, we must consider the origin of the bunch modulating instability. Fast Brunel electrons penetrate into the plasma mirror and induce a return current composed of counter-streaming electrons and ions flowing along the surface. Assuming that the drift velocity of the return electrons is much larger than ions $v_{de} \gg v_{di}$, the ion velocity may be neglected and, following the usual treatment of two-stream instability, we obtain the well-known Buneman instability (BI) dispersion relation for a small perturbation \citep{buneman1959dissipation}, 
\begin{equation}
\label{eq:2}
    \frac{\omega_{pe}^{2}}{(\omega - kv_{de})^{2}} + \frac{\omega_{pi}^{2}}{\omega^{2}} = 1,
\end{equation}
where $\omega_{pi} = \left(Z^{2}e^{2}n_{i}/m_{i}\epsilon_{0}\right)^{1/2}$ is the ion plasma frequency. {\color{black}For an unstable growing mode, the mirror surface, and therefore} the Brunel electrons, will be modulated according to the instability wavelength as illustrated in Fig. \ref{fig:1}c. Solving Eq. \ref{eq:2} for the complex frequency $\omega \rightarrow \omega + i\Gamma$, we obtain from the imaginary part the maximum growth rate of the Buneman instability as 
\begin{equation}\label{eq:3}
    \Gamma _{m} = \omega_{pe}\frac{\sqrt{3}}{2} \left(\frac{Z^{2}m_{e}}{2m_{i}}\right)^{1/3},
\end{equation}
which is approximately valid for all modes in the band $\vert kv_{de} - \omega_{p}\vert \lesssim (3/2)(\omega_{pe}\omega_{pi}^{2})^{1/3}$ \citep{buneman1959dissipation}. {\color{black}The condition for RIME to occur therefore requires laser pulse duration to be larger than $\approx 1/\Gamma_{m}$}. This straightforward result allows us to see that the plasma surface instability e-folding time can be of the order of a single laser cycle, since $\Gamma _{m}\approx \omega_{0}$ for solid density plasma, $n_{e} \approx 10^{3}n_{c}$, which means that the plasma surface instability can fully manifest already within a few laser cycles.

To study the process in detail, we have performed multi-dimensional particle-in-cell (PIC) simulations (See Supplementary Material \citep{lamac2022supp} for simulation details). A P-polarized laser pulse defined with pulse duration $\tau_{fwhm} = 30$ fs, laser wavelength $\lambda_{0} = 1\text{ }\mu$m and peak intensity $I_{0} = 10^{22}\text{ W/cm}^{2}$ corresponding to a normalized laser amplitude $a_{0} = 85.5$, was focused with an incidence angle $\theta = 45^{\circ}$ to a spot size with waist radius $w_{0} = 2\text{ }\mu$m upon a plasma mirror composed of uniformly overlapping electrons and ions with matching density $ n_{i}/n_{c} = n_{e}/n_{c} = 1000$ and temperature $k_{B}T = 100\text{ eV}$. {\color{black}Figure \ref{fig:2} shows the analysis of plasma surface instability observed in the PIC simulation. Fig. \ref{fig:2}a shows the temporal evolution of the surface-parallel ion current, revealing an unstable plasma wave propagating in the direction of electron return flow with phase velocity $v_{p}/c \approx 0.01$. This corresponds to the drift velocity of the electron return current, since from the current neutrality condition $v_{de}/c \approx a_{0}n_{c}/n_{e} \approx 0.01$ at $t \approx 45 $ fs when the amplitude of the incident laser is $a_{0} \approx 10$. Fig. \ref{fig:2}b presents the time evolution of the instability, showing the instability saturation and clear separation between the linear and non-linear phases occuring at $t \approx 55$ fs. We note that the maximum instantaneous growth rate matches the prediction given by Eq. \ref{eq:3}. Fig. \ref{fig:2}c shows the numerical dispersion of the instability overlaid with the real part of the solution of Eq. \ref{eq:2}, showing a match between the PIC and Buneman unstable modes.

Figure \ref{fig:3} presents a detailed view of the RIME origin. In Fig. \ref{fig:3}a, Brunel electrons can be seen penetrating into the bulk in the direction of laser propagation at twice the laser frequency, which is due to the relativistic $\mathbf{j}\times\mathbf{B}$ Lorentz force term. This leads to the growth of unstable return current flowing along the periphery seen in Fig. \ref{fig:3}b. As the unstable plasma wave breaks, oscillating electron nanobunches are accelerated by the laser across the surface within the focal region to relativistic velocities. This is shown with the electron current components in Figs. \ref{fig:3}c, d. The transverse component highlights the transverse oscillatory turning points of the individual nanobunches occuring at $j_{\perp, e} = 0$. On the other hand, it is at these points where longitudinal velocity is largest, $v_{\parallel} \approx c$, which leads to relativistic beaming of radiation along the surface (See Supplementary Material \citep{lamac2022supp} for a movie of the process). This results in the loss of reflected wave coherence and the emergence of RIME XUV bursts seen in the perpendicular component of the magnetic field $B_{z}$ shown in Fig. \ref{fig:3}f. Additionally, the magnetic field reveals that the instability region is extremely magnetized due to the return current, with amplitude of the order of the incident laser. This leads to enhanced confinement of relativistic electrons towards the plasma surface \citep{nakamura2004surface}. The electric field component $E_{y}$ in Fig. \ref{fig:3}e shows the incident P-polarized laser interfering with the intense RIME bursts.}

\begin{figure*}
\includegraphics[scale = 0.345]{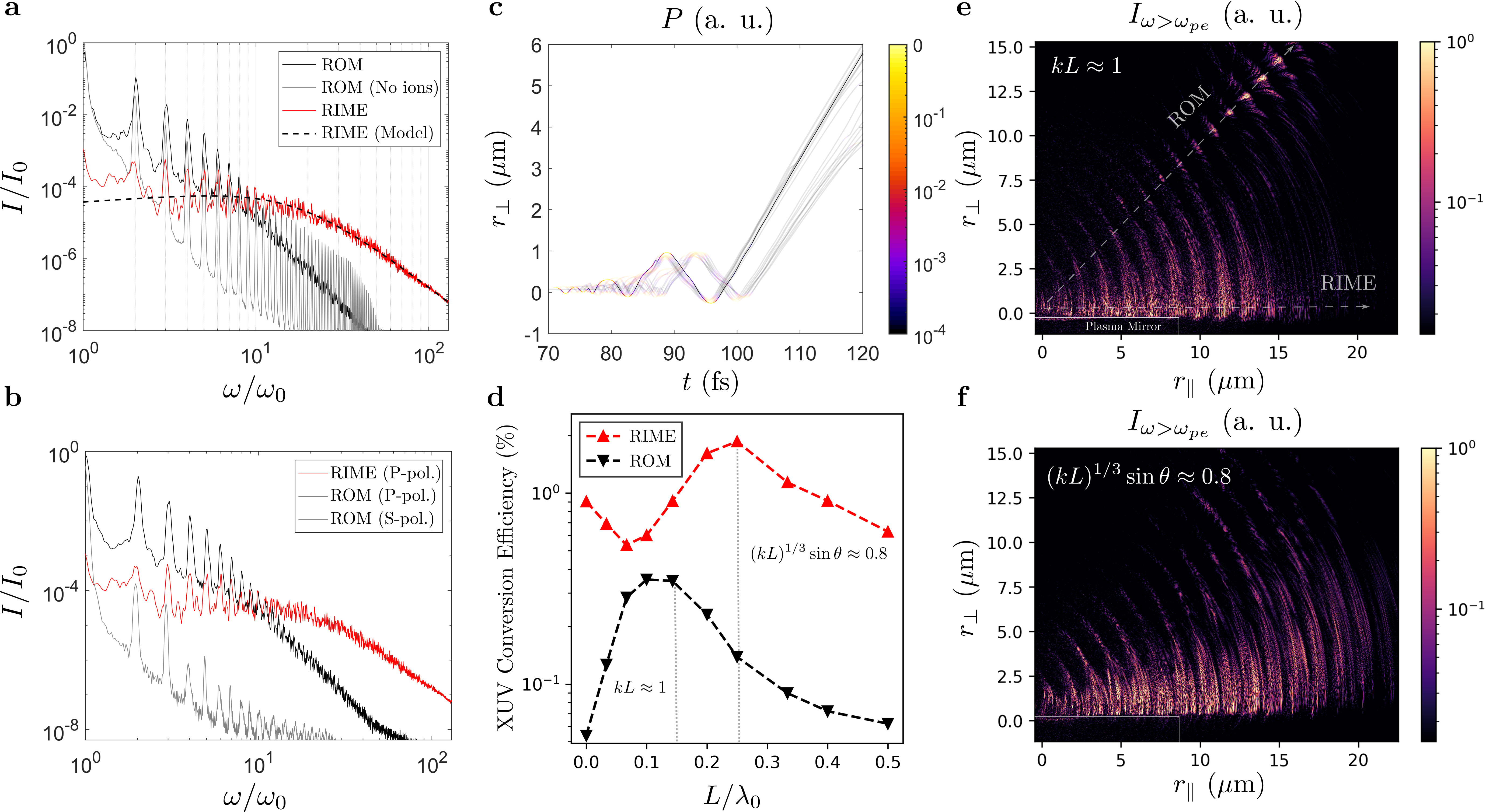}
\caption{\textbf{Radiation properties of RIME.} {\color{black}a) Intensity spectra with the analytical model given by Eq. \ref{eq:1}. b) Intensity spectra for S and P laser polarizations. c)  Instantaneous power of radiation emitted along the surface calculated for two electron bunches oscillating above the surface ($r_{\perp} = 0$) with $\gamma \approx a_{0}$. d) Laser-to-XUV energy conversion efficiency dependence on pre-plasma scale length $L$. e) Spatial distribution of XUV intensity for the ROM-optimal and f) the RIME-optimal pre-plasma.}}
\label{fig:4}
\end{figure*}

Radiation characteristics of RIME are presented in Fig. \ref{fig:4}. Trajectories of two characteristic electron bunches with $\gamma \approx a_{0}$ are presented in Fig. \ref{fig:4}c. {\color{black}Calculated radiation power confirms multiple bursts emitted along the bunch trajectory. The intensity spectra of RIME and the wave reflected due to the relativistic oscillations of the mirror (ROM) are shown in Fig. \ref{fig:4}a. The spatio-temporal coherence of the reflected wave is reduced with the onset of RIME, leading to larger XUV conversion efficiency for the emission along the mirror surface. To evaluate coherent enhancement of radiation
due to electron bunching, we have performed another PIC simulation with immobile ions. The simulation confirmed that in this case BI and RIME are not present. In this scenario, the coherence of high harmonics composing the reflected wave is significantly improved, as shown in Fig. \ref{fig:4}a. However, compared to the RIME spectrum, the efficiency significantly drops for $\omega/\omega_{0} > 10$, which can be explained by the nanobunching effect occuring in the realistic simulation, since the individual bunch sizes can be smaller than $\lambda_{pe} = \lambda_{0}/31.6 \approx 31 \text{ nm}$ as shown in Fig. \ref{fig:3}c, d. The continuous RIME spectrum is explained by the fact that the peak laser-plasma interaction, and therefore strongest emission, occurs deep into the non-linear phase of the instability evolution at $t = 80$ fs, as seen in Fig. \ref{fig:2}b. This leads to non-periodic trains of electron nanobunches individually producing broadband XUV bursts that add up to a broad, continuous spectrum. We point out that the bunching effect also enhances the low-order harmonics composing the reflected wave, which can be seen when comparing the two ROM spectra shown in Fig. \ref{fig:4}a. This is because the distances between the individual bunches within a single laser oscillation are smaller than the wavelengths of the low-order harmonics, which can be seen in Figs. \ref{fig:3}e, f.} To compare the RIME spectrum with the analytical result given by Eq. \ref{eq:1}, we proceed to estimate the bunch modulation frequency. The average value of electron drift velocity inside the return current observed in the PIC simulation is $v_{de}/c \approx 0.18$ at the peak interaction time $t = 80$ fs (Fig. \ref{fig:3}d). The bunch modulation frequency due to BI can be therefore estimated as $\omega_{b} = c/L_{b} \approx 2c/\lambda_{B} \approx (c\omega_{p}/v_{de}\pi) \approx 18\omega_{0}$, corresponding to an electron bunch length of $L_{b} \approx 88\text{ nm}$ and instability wavelength $\lambda_{B} \approx 0.18\text{ }\mu\text{m}$, which is in agreement with the features seen in Fig. \ref{fig:3}c, d. Considering such modulation frequency, RIME spectrum as given by Eq. \ref{eq:1}, with energy of laser-accelerated electrons given $\gamma\approx a_{0} = 85.5$, is presented in Fig. \ref{fig:4}a, showing an excellent agreement with the PIC result. In Fig. \ref{fig:4}b we show the dependence of emitted radiation on laser polarization. When the incident laser is S-polarized, BI and RIME are not present and only ROM-reflected wave remains with less efficient high-harmonic generation, which confirms a well-known property of ROM \citep{teubner2009high}.

Previous studies have demonstrated that the presence of pre-plasma with scale length satisfying $kL \approx 1$ can optimize ROM conversion efficiency \citep{kahaly2013direct, dollar2013scaling}, where $k = \omega_{0} /c$ is the laser wave number number and $L$ is the pre-plasma scale length.  {\color{black}This optimum comes from the balance of the strong electrostatic field due to the space-charge distribution limiting electron oscillations when $kL < 1$, and the growth of plasma waves leading to reflected wave coherence loss when $kL > 1$ \citep{ginzburg1966inplasma, gibbon2005short}. In the case when $kL > 1$, the plasma wave growth is resonant for a P-polarized laser incident upon inhomogeneous plasma with a scale length satisfying \citep{deniov1957resabs, ginzburg1966inplasma}
\begin{equation}\label{eq4}
(kL)^{1/3}\sin \theta \approx 0.8,
\end{equation}
where $\theta $ is the laser incidence angle. To investigate the effect of non-zero pre-plasma scale lengths on RIME, we have conducted PIC simulations in the range $0 \leq kL \leq \pi$. The results are summarized in Fig. \ref{fig:4}d, where we show the conversion efficiency of laser energy to XUV wavelengths $\lambda \leq 100$ nm. Our numerical results match the experimentally discovered ROM optimum found in \citep{kahaly2013direct, dollar2013scaling}, which occurs when $kL \approx 1$. For such scale length, the total reflectivity reaches up to $60 \%$, with around $0.4 \%$ of laser energy converted to XUV. The results for RIME show high XUV efficiency, but more intricate dependence. First, as discussed above, the coherent enhancement due to nanobunching yields larger XUV conversion efficiency by up to an order-of-magnitude. Second, the XUV yields are anti-correlated. This is due to the fact that the coherent enhancement of RIME results in the loss of coherence in the reflected wave and vice versa.  Finally, the RIME XUV yield reveals an optimal scale length for the simulation with $L/\lambda_{0} = 0.25$, where XUV conversion efficiency grows up to $2 \%$. To compare to conventional XUV sources, the widely-used gas HHG source operates at XUV energy conversion efficiency of $\approx 10^{-5}$ \citep{nefedova2017development, heyl2016scale}, with maximum incident laser intensity limited to $\sim 10^{15}\text{ W.cm}^{-2}$. 

\begin{figure}
\includegraphics[scale = 0.53]{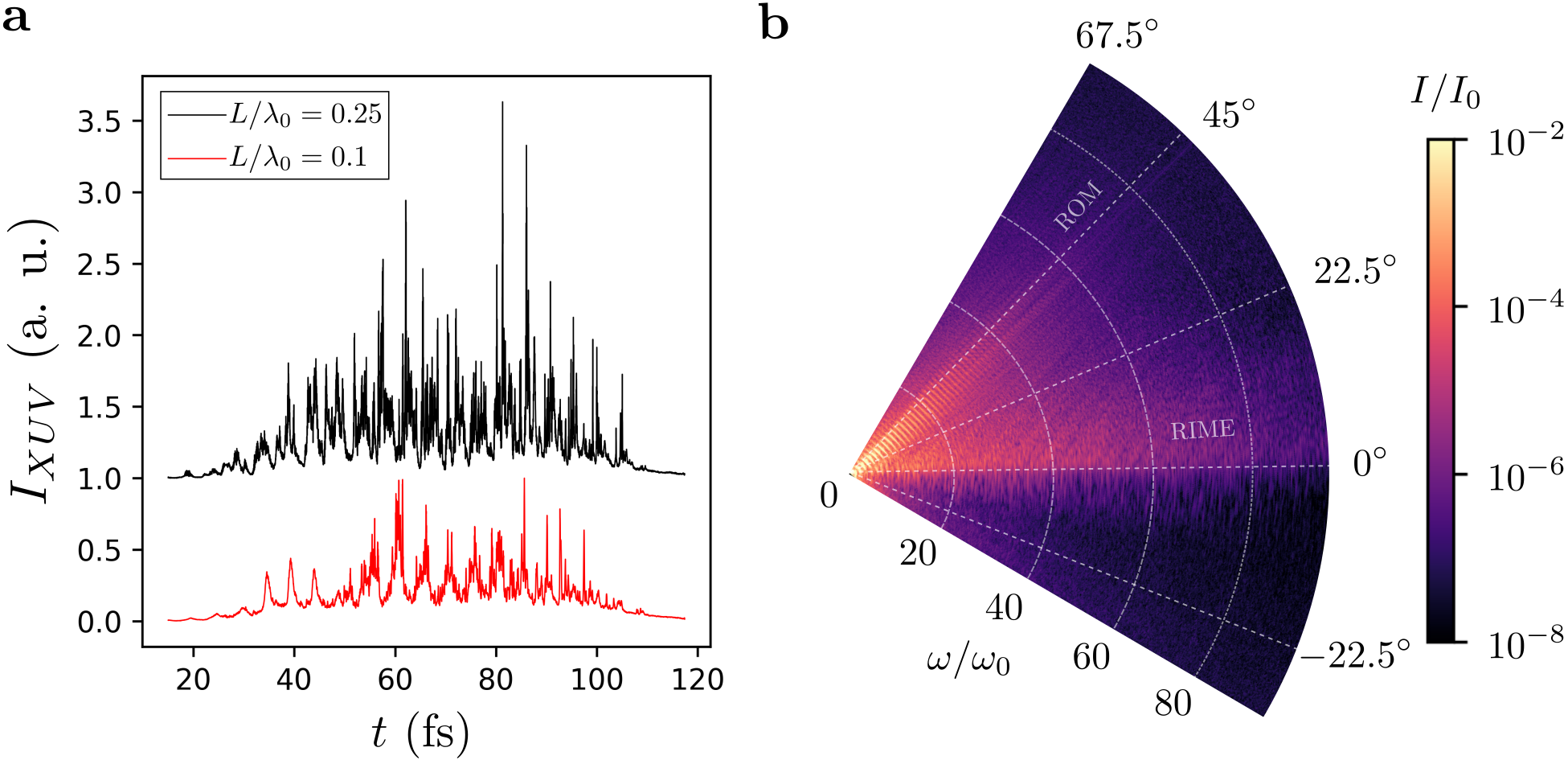}
\caption{{\color{black}a) Temporal profile of XUV radiation for the case of RIME (black) and ROM (red) optimal scale length, offset for clarity. b) Angular distribution of spectral intensity for ROM-optimal scale length $L/\lambda_{0} = 0.1$. Colormap is saturated for higher contrast in the XUV region.}} 
\label{fig:5}
\end{figure}

For the simulated incidence angle of $45^{\circ}$, the condition for resonant plasma wave growth given by Eq. \ref{eq4} gives the scale length as $L/\lambda_{0} = 0.23$, an excellent match which shows that the origin of the RIME optimum is due to the resonant growth of large-amplitude plasma waves which facilitate faster wave-breaking of instability-modulated electrons into the laser field. The XUV intensity distribution for the two optimal cases are shown in Figs. \ref{fig:4}e, f. In the case of $kL \approx 1$, the laser is initially reflected coherently by ROM. As the unstable plasma wave saturates, the spatio-temporal XUV coherence is reduced and coherent XUV bursts propagating along the mirror surface emerge. In the case of resonant plasma wave growth, $(kL)^{1/3}\sin \theta \approx 0.8$, the reflected wave loses its spatio-temporal coherence almost immediatelly and most of the XUV radiation is increasingly collimated towards the surface as the laser intensity ramps up. 

The angular distribution of RIME is given by the emission angle of laser-accelerated relativistic electrons given by $\tan \theta = (2/(\gamma - 1))^{1/2}$ \citep{gibbon2005short} . For the electrons oscillating in the peak laser amplitude $\gamma \approx a_{0}$, the emission angle is therefore $\theta \approx 8.7 ^{\circ} = 150$ mrad, which is in reasonable agreement with the XUV intensity distributions shown in Figs. \ref{fig:4}e, f, and the angular distribution of RIME for the case of $kL \approx 1$ shown in Fig. \ref{fig:5}b. The angular distribution shows clear separation between the high-frequency radiation emitted by ROM and RIME, which is also visible in the corresponding XUV intensity distribution shown in Fig. \ref{fig:4}e. The temporal profile of XUV radiation for the two optimal cases is shown in Fig. \ref{fig:5}a. In correspondence with Fig. \ref{fig:4}e, the radiation pulse of the ROM-optimal case $L/\lambda_{0} = 0.1$ shows initially coherent and smooth reflected wave profile at $t < 55$ fs, losing its coherence at $t = 55$ fs, and then transitioning to the RIME profile modulated by coherent attosecond bursts at $t > 55$ fs. The case of RIME-optimal scale length $L/\lambda_{0} = 0.25$ shows that alredy at the beginning of interaction, $t < 55$ fs, the XUV pulse is modulated with the RIME bursts in accordance with Fig. \ref{fig:4}f. As the laser peak arrives at $t = 80$ fs, coherently-enhanced giant attosecond XUV pulses emerge due to the strong electron nanobunching from the resonant plasma wave growth.

In conclusion, a new high-yield mechanism of coherent XUV light generation called RIME was discovered to occur during the interaction of relativistically intense {\color{black}P-polarized} laser pulse with overdense plasma {\color{black}satisfying $1 \leq a_{0} < n_{e}/n_{c}$}. The radiation emission is directionally anomalous, propagating parallel to the mirror surface. With numerical and analytical methods, we have found that RIME occurs due to laser-driven oscillations of electron nanobunches which emerge from a return current instability induced by collisionless absorption of the incident laser. Furthermore, by varying the pre-plasma scale length, we have found that RIME is anti-correlated with the reflected high-harmonics, showing the possibility to tune between two directionally-separated XUV radiation sources in a single experimental configuration. Finally, we have found an optimal pre-plasma scale length for RIME corresponding to resonant growth of plasma waves at the mirror surface. The optimal XUV conversion efficiency of RIME reaches percent-level, presenting a new approach for future experiments aiming to produce high-yield coherent XUV radiation. 

}

\section*{Acknowledgements}

This work was supported by the project ADONIS (CZ.02.1.01/0.0/0.0/16\_019/0000789). 

\nocite{*}

\bibliography{apssamp}

\providecommand{\noopsort}[1]{}\providecommand{\singleletter}[1]{#1}%
\begin{thebibliography}{31}%
\makeatletter
\providecommand \@ifxundefined [1]{%
 \@ifx{#1\undefined}
}%
\providecommand \@ifnum [1]{%
 \ifnum #1\expandafter \@firstoftwo
 \else \expandafter \@secondoftwo
 \fi
}%
\providecommand \@ifx [1]{%
 \ifx #1\expandafter \@firstoftwo
 \else \expandafter \@secondoftwo
 \fi
}%
\providecommand \natexlab [1]{#1}%
\providecommand \enquote  [1]{``#1''}%
\providecommand \bibnamefont  [1]{#1}%
\providecommand \bibfnamefont [1]{#1}%
\providecommand \citenamefont [1]{#1}%
\providecommand \href@noop [0]{\@secondoftwo}%
\providecommand \href [0]{\begingroup \@sanitize@url \@href}%
\providecommand \@href[1]{\@@startlink{#1}\@@href}%
\providecommand \@@href[1]{\endgroup#1\@@endlink}%
\providecommand \@sanitize@url [0]{\catcode `\\12\catcode `\$12\catcode
  `\&12\catcode `\#12\catcode `\^12\catcode `\_12\catcode `\%12\relax}%
\providecommand \@@startlink[1]{}%
\providecommand \@@endlink[0]{}%
\providecommand \url  [0]{\begingroup\@sanitize@url \@url }%
\providecommand \@url [1]{\endgroup\@href {#1}{\urlprefix }}%
\providecommand \urlprefix  [0]{URL }%
\providecommand \Eprint [0]{\href }%
\providecommand \doibase [0]{https://doi.org/}%
\providecommand \selectlanguage [0]{\@gobble}%
\providecommand \bibinfo  [0]{\@secondoftwo}%
\providecommand \bibfield  [0]{\@secondoftwo}%
\providecommand \translation [1]{[#1]}%
\providecommand \BibitemOpen [0]{}%
\providecommand \bibitemStop [0]{}%
\providecommand \bibitemNoStop [0]{.\EOS\space}%
\providecommand \EOS [0]{\spacefactor3000\relax}%
\providecommand \BibitemShut  [1]{\csname bibitem#1\endcsname}%
\let\auto@bib@innerbib\@empty
\bibitem [{\citenamefont {Mori}\ \emph {et~al.}(1993)\citenamefont {Mori},
  \citenamefont {Decker},\ and\ \citenamefont
  {Leemans}}]{mori1993relativistic}%
  \BibitemOpen
  \bibfield  {author} {\bibinfo {author} {\bibfnamefont {W.}~\bibnamefont
  {Mori}}, \bibinfo {author} {\bibfnamefont {C.}~\bibnamefont {Decker}},\ and\
  \bibinfo {author} {\bibfnamefont {W.}~\bibnamefont {Leemans}},\ }\bibfield
  {title} {\bibinfo {title} {Relativistic harmonic content of nonlinear
  electromagnetic waves in underdense plasmas},\ }\href@noop {} {\bibfield
  {journal} {\bibinfo  {journal} {IEEE transactions on plasma science}\
  }\textbf {\bibinfo {volume} {21}},\ \bibinfo {pages} {110} (\bibinfo {year}
  {1993})}\BibitemShut {NoStop}%
\bibitem [{\citenamefont {Teubner}\ and\ \citenamefont
  {Gibbon}(2009)}]{teubner2009high}%
  \BibitemOpen
  \bibfield  {author} {\bibinfo {author} {\bibfnamefont {U.}~\bibnamefont
  {Teubner}}\ and\ \bibinfo {author} {\bibfnamefont {P.}~\bibnamefont
  {Gibbon}},\ }\bibfield  {title} {\bibinfo {title} {High-order harmonics from
  laser-irradiated plasma surfaces},\ }\href@noop {} {\bibfield  {journal}
  {\bibinfo  {journal} {Reviews of Modern Physics}\ }\textbf {\bibinfo {volume}
  {81}},\ \bibinfo {pages} {445} (\bibinfo {year} {2009})}\BibitemShut
  {NoStop}%
\bibitem [{\citenamefont {Bulanov}\ \emph {et~al.}(1994)\citenamefont
  {Bulanov}, \citenamefont {Naumova},\ and\ \citenamefont
  {Pegoraro}}]{bulanov1994interaction}%
  \BibitemOpen
  \bibfield  {author} {\bibinfo {author} {\bibfnamefont {S.~V.}\ \bibnamefont
  {Bulanov}}, \bibinfo {author} {\bibfnamefont {N.}~\bibnamefont {Naumova}},\
  and\ \bibinfo {author} {\bibfnamefont {F.}~\bibnamefont {Pegoraro}},\
  }\bibfield  {title} {\bibinfo {title} {Interaction of an ultrashort,
  relativistically strong laser pulse with an overdense plasma},\ }\href@noop
  {} {\bibfield  {journal} {\bibinfo  {journal} {Physics of Plasmas}\ }\textbf
  {\bibinfo {volume} {1}},\ \bibinfo {pages} {745} (\bibinfo {year}
  {1994})}\BibitemShut {NoStop}%
\bibitem [{\citenamefont {Vincenti}\ \emph {et~al.}(2014)\citenamefont
  {Vincenti}, \citenamefont {Monchoc{\'e}}, \citenamefont {Kahaly},
  \citenamefont {Bonnaud}, \citenamefont {Martin},\ and\ \citenamefont
  {Qu{\'e}r{\'e}}}]{vincenti2014optical}%
  \BibitemOpen
  \bibfield  {author} {\bibinfo {author} {\bibfnamefont {H.}~\bibnamefont
  {Vincenti}}, \bibinfo {author} {\bibfnamefont {S.}~\bibnamefont
  {Monchoc{\'e}}}, \bibinfo {author} {\bibfnamefont {S.}~\bibnamefont
  {Kahaly}}, \bibinfo {author} {\bibfnamefont {G.}~\bibnamefont {Bonnaud}},
  \bibinfo {author} {\bibfnamefont {P.}~\bibnamefont {Martin}},\ and\ \bibinfo
  {author} {\bibfnamefont {F.}~\bibnamefont {Qu{\'e}r{\'e}}},\ }\bibfield
  {title} {\bibinfo {title} {Optical properties of relativistic plasma
  mirrors},\ }\href@noop {} {\bibfield  {journal} {\bibinfo  {journal} {Nature
  communications}\ }\textbf {\bibinfo {volume} {5}},\ \bibinfo {pages} {3403}
  (\bibinfo {year} {2014})}\BibitemShut {NoStop}%
\bibitem [{\citenamefont {Pirozhkov}\ \emph {et~al.}(2012)\citenamefont
  {Pirozhkov}, \citenamefont {Kando}, \citenamefont {Esirkepov}, \citenamefont
  {Gallegos}, \citenamefont {Ahmed}, \citenamefont {Ragozin}, \citenamefont
  {Faenov}, \citenamefont {Pikuz}, \citenamefont {Kawachi}, \citenamefont
  {Sagisaka} \emph {et~al.}}]{pirozhkov2012soft}%
  \BibitemOpen
  \bibfield  {author} {\bibinfo {author} {\bibfnamefont {A.}~\bibnamefont
  {Pirozhkov}}, \bibinfo {author} {\bibfnamefont {M.}~\bibnamefont {Kando}},
  \bibinfo {author} {\bibfnamefont {T.~Z.}\ \bibnamefont {Esirkepov}}, \bibinfo
  {author} {\bibfnamefont {P.}~\bibnamefont {Gallegos}}, \bibinfo {author}
  {\bibfnamefont {H.}~\bibnamefont {Ahmed}}, \bibinfo {author} {\bibfnamefont
  {E.}~\bibnamefont {Ragozin}}, \bibinfo {author} {\bibfnamefont {A.~Y.}\
  \bibnamefont {Faenov}}, \bibinfo {author} {\bibfnamefont {T.}~\bibnamefont
  {Pikuz}}, \bibinfo {author} {\bibfnamefont {T.}~\bibnamefont {Kawachi}},
  \bibinfo {author} {\bibfnamefont {A.}~\bibnamefont {Sagisaka}}, \emph
  {et~al.},\ }\bibfield  {title} {\bibinfo {title} {Soft-x-ray harmonic comb
  from relativistic electron spikes},\ }\href@noop {} {\bibfield  {journal}
  {\bibinfo  {journal} {Physical Review Letters}\ }\textbf {\bibinfo {volume}
  {108}},\ \bibinfo {pages} {135004} (\bibinfo {year} {2012})}\BibitemShut
  {NoStop}%
\bibitem [{\citenamefont {Dromey}\ \emph {et~al.}(2009)\citenamefont {Dromey},
  \citenamefont {Adams}, \citenamefont {H{\"o}rlein}, \citenamefont {Nomura},
  \citenamefont {Rykovanov}, \citenamefont {Carroll}, \citenamefont {Foster},
  \citenamefont {Kar}, \citenamefont {Markey}, \citenamefont {McKenna} \emph
  {et~al.}}]{dromey2009diffraction}%
  \BibitemOpen
  \bibfield  {author} {\bibinfo {author} {\bibfnamefont {B.}~\bibnamefont
  {Dromey}}, \bibinfo {author} {\bibfnamefont {D.}~\bibnamefont {Adams}},
  \bibinfo {author} {\bibfnamefont {R.}~\bibnamefont {H{\"o}rlein}}, \bibinfo
  {author} {\bibfnamefont {Y.}~\bibnamefont {Nomura}}, \bibinfo {author}
  {\bibfnamefont {S.}~\bibnamefont {Rykovanov}}, \bibinfo {author}
  {\bibfnamefont {D.}~\bibnamefont {Carroll}}, \bibinfo {author} {\bibfnamefont
  {P.}~\bibnamefont {Foster}}, \bibinfo {author} {\bibfnamefont
  {S.}~\bibnamefont {Kar}}, \bibinfo {author} {\bibfnamefont {K.}~\bibnamefont
  {Markey}}, \bibinfo {author} {\bibfnamefont {P.}~\bibnamefont {McKenna}},
  \emph {et~al.},\ }\bibfield  {title} {\bibinfo {title} {Diffraction-limited
  performance and focusing of high harmonics from relativistic plasmas},\
  }\href@noop {} {\bibfield  {journal} {\bibinfo  {journal} {Nature Physics}\
  }\textbf {\bibinfo {volume} {5}},\ \bibinfo {pages} {146} (\bibinfo {year}
  {2009})}\BibitemShut {NoStop}%
\bibitem [{\citenamefont {Yeung}\ \emph {et~al.}(2014)\citenamefont {Yeung},
  \citenamefont {Dromey}, \citenamefont {Cousens}, \citenamefont {Dzelzainis},
  \citenamefont {Kiefer}, \citenamefont {Schreiber}, \citenamefont {Bin},
  \citenamefont {Ma}, \citenamefont {Kreuzer}, \citenamefont {Meyer-ter Vehn}
  \emph {et~al.}}]{yeung2014dependence}%
  \BibitemOpen
  \bibfield  {author} {\bibinfo {author} {\bibfnamefont {M.}~\bibnamefont
  {Yeung}}, \bibinfo {author} {\bibfnamefont {B.}~\bibnamefont {Dromey}},
  \bibinfo {author} {\bibfnamefont {S.}~\bibnamefont {Cousens}}, \bibinfo
  {author} {\bibfnamefont {T.}~\bibnamefont {Dzelzainis}}, \bibinfo {author}
  {\bibfnamefont {D.}~\bibnamefont {Kiefer}}, \bibinfo {author} {\bibfnamefont
  {J.}~\bibnamefont {Schreiber}}, \bibinfo {author} {\bibfnamefont
  {J.}~\bibnamefont {Bin}}, \bibinfo {author} {\bibfnamefont {W.}~\bibnamefont
  {Ma}}, \bibinfo {author} {\bibfnamefont {C.}~\bibnamefont {Kreuzer}},
  \bibinfo {author} {\bibfnamefont {J.}~\bibnamefont {Meyer-ter Vehn}}, \emph
  {et~al.},\ }\bibfield  {title} {\bibinfo {title} {Dependence of laser-driven
  coherent synchrotron emission efficiency on pulse ellipticity and
  implications for polarization gating},\ }\href@noop {} {\bibfield  {journal}
  {\bibinfo  {journal} {Physical Review Letters}\ }\textbf {\bibinfo {volume}
  {112}},\ \bibinfo {pages} {123902} (\bibinfo {year} {2014})}\BibitemShut
  {NoStop}%
\bibitem [{\citenamefont {Heissler}\ \emph {et~al.}(2012)\citenamefont
  {Heissler}, \citenamefont {H{\"o}rlein}, \citenamefont {Mikhailova},
  \citenamefont {Waldecker}, \citenamefont {Tzallas}, \citenamefont {Buck},
  \citenamefont {Schmid}, \citenamefont {Sears}, \citenamefont {Krausz},
  \citenamefont {Veisz} \emph {et~al.}}]{heissler2012few}%
  \BibitemOpen
  \bibfield  {author} {\bibinfo {author} {\bibfnamefont {P.}~\bibnamefont
  {Heissler}}, \bibinfo {author} {\bibfnamefont {R.}~\bibnamefont
  {H{\"o}rlein}}, \bibinfo {author} {\bibfnamefont {J.~M.}\ \bibnamefont
  {Mikhailova}}, \bibinfo {author} {\bibfnamefont {L.}~\bibnamefont
  {Waldecker}}, \bibinfo {author} {\bibfnamefont {P.}~\bibnamefont {Tzallas}},
  \bibinfo {author} {\bibfnamefont {A.}~\bibnamefont {Buck}}, \bibinfo {author}
  {\bibfnamefont {K.}~\bibnamefont {Schmid}}, \bibinfo {author} {\bibfnamefont
  {C.}~\bibnamefont {Sears}}, \bibinfo {author} {\bibfnamefont
  {F.}~\bibnamefont {Krausz}}, \bibinfo {author} {\bibfnamefont
  {L.}~\bibnamefont {Veisz}}, \emph {et~al.},\ }\bibfield  {title} {\bibinfo
  {title} {Few-cycle driven relativistically oscillating plasma mirrors: a
  source of intense isolated attosecond pulses},\ }\href@noop {} {\bibfield
  {journal} {\bibinfo  {journal} {Physical Review Letters}\ }\textbf {\bibinfo
  {volume} {108}},\ \bibinfo {pages} {235003} (\bibinfo {year}
  {2012})}\BibitemShut {NoStop}%
\bibitem [{\citenamefont {Naumova}\ \emph {et~al.}(2004)\citenamefont
  {Naumova}, \citenamefont {Nees}, \citenamefont {Sokolov}, \citenamefont
  {Hou},\ and\ \citenamefont {Mourou}}]{naumova2004relativistic}%
  \BibitemOpen
  \bibfield  {author} {\bibinfo {author} {\bibfnamefont {N.}~\bibnamefont
  {Naumova}}, \bibinfo {author} {\bibfnamefont {J.}~\bibnamefont {Nees}},
  \bibinfo {author} {\bibfnamefont {I.}~\bibnamefont {Sokolov}}, \bibinfo
  {author} {\bibfnamefont {B.}~\bibnamefont {Hou}},\ and\ \bibinfo {author}
  {\bibfnamefont {G.}~\bibnamefont {Mourou}},\ }\bibfield  {title} {\bibinfo
  {title} {Relativistic generation of isolated attosecond pulses in a $\lambda$
  3 focal volume},\ }\href@noop {} {\bibfield  {journal} {\bibinfo  {journal}
  {Physical Review Letters}\ }\textbf {\bibinfo {volume} {92}},\ \bibinfo
  {pages} {063902} (\bibinfo {year} {2004})}\BibitemShut {NoStop}%
\bibitem [{\citenamefont {Wheeler}\ \emph {et~al.}(2012)\citenamefont
  {Wheeler}, \citenamefont {Borot}, \citenamefont {Monchoc{\'e}}, \citenamefont
  {Vincenti}, \citenamefont {Ricci}, \citenamefont {Malvache}, \citenamefont
  {Lopez-Martens},\ and\ \citenamefont
  {Qu{\'e}r{\'e}}}]{wheeler2012attosecond}%
  \BibitemOpen
  \bibfield  {author} {\bibinfo {author} {\bibfnamefont {J.~A.}\ \bibnamefont
  {Wheeler}}, \bibinfo {author} {\bibfnamefont {A.}~\bibnamefont {Borot}},
  \bibinfo {author} {\bibfnamefont {S.}~\bibnamefont {Monchoc{\'e}}}, \bibinfo
  {author} {\bibfnamefont {H.}~\bibnamefont {Vincenti}}, \bibinfo {author}
  {\bibfnamefont {A.}~\bibnamefont {Ricci}}, \bibinfo {author} {\bibfnamefont
  {A.}~\bibnamefont {Malvache}}, \bibinfo {author} {\bibfnamefont
  {R.}~\bibnamefont {Lopez-Martens}},\ and\ \bibinfo {author} {\bibfnamefont
  {F.}~\bibnamefont {Qu{\'e}r{\'e}}},\ }\bibfield  {title} {\bibinfo {title}
  {Attosecond lighthouses from plasma mirrors},\ }\href@noop {} {\bibfield
  {journal} {\bibinfo  {journal} {Nature Photonics}\ }\textbf {\bibinfo
  {volume} {6}},\ \bibinfo {pages} {829} (\bibinfo {year} {2012})}\BibitemShut
  {NoStop}%
\bibitem [{\citenamefont {Smirnova}\ \emph {et~al.}(2009)\citenamefont
  {Smirnova}, \citenamefont {Mairesse}, \citenamefont {Patchkovskii},
  \citenamefont {Dudovich}, \citenamefont {Villeneuve}, \citenamefont
  {Corkum},\ and\ \citenamefont {Ivanov}}]{smirnova2009high}%
  \BibitemOpen
  \bibfield  {author} {\bibinfo {author} {\bibfnamefont {O.}~\bibnamefont
  {Smirnova}}, \bibinfo {author} {\bibfnamefont {Y.}~\bibnamefont {Mairesse}},
  \bibinfo {author} {\bibfnamefont {S.}~\bibnamefont {Patchkovskii}}, \bibinfo
  {author} {\bibfnamefont {N.}~\bibnamefont {Dudovich}}, \bibinfo {author}
  {\bibfnamefont {D.}~\bibnamefont {Villeneuve}}, \bibinfo {author}
  {\bibfnamefont {P.}~\bibnamefont {Corkum}},\ and\ \bibinfo {author}
  {\bibfnamefont {M.~Y.}\ \bibnamefont {Ivanov}},\ }\bibfield  {title}
  {\bibinfo {title} {High harmonic interferometry of multi-electron dynamics in
  molecules},\ }\href@noop {} {\bibfield  {journal} {\bibinfo  {journal}
  {Nature}\ }\textbf {\bibinfo {volume} {460}},\ \bibinfo {pages} {972}
  (\bibinfo {year} {2009})}\BibitemShut {NoStop}%
\bibitem [{\citenamefont {Neutze}\ \emph {et~al.}(2000)\citenamefont {Neutze},
  \citenamefont {Wouts}, \citenamefont {Van~der Spoel}, \citenamefont
  {Weckert},\ and\ \citenamefont {Hajdu}}]{neutze2000potential}%
  \BibitemOpen
  \bibfield  {author} {\bibinfo {author} {\bibfnamefont {R.}~\bibnamefont
  {Neutze}}, \bibinfo {author} {\bibfnamefont {R.}~\bibnamefont {Wouts}},
  \bibinfo {author} {\bibfnamefont {D.}~\bibnamefont {Van~der Spoel}}, \bibinfo
  {author} {\bibfnamefont {E.}~\bibnamefont {Weckert}},\ and\ \bibinfo {author}
  {\bibfnamefont {J.}~\bibnamefont {Hajdu}},\ }\bibfield  {title} {\bibinfo
  {title} {Potential for biomolecular imaging with femtosecond x-ray pulses},\
  }\href@noop {} {\bibfield  {journal} {\bibinfo  {journal} {Nature}\ }\textbf
  {\bibinfo {volume} {406}},\ \bibinfo {pages} {752} (\bibinfo {year}
  {2000})}\BibitemShut {NoStop}%
\bibitem [{\citenamefont {Krausz}\ and\ \citenamefont
  {Ivanov}(2009)}]{krausz2009attosecond}%
  \BibitemOpen
  \bibfield  {author} {\bibinfo {author} {\bibfnamefont {F.}~\bibnamefont
  {Krausz}}\ and\ \bibinfo {author} {\bibfnamefont {M.}~\bibnamefont
  {Ivanov}},\ }\bibfield  {title} {\bibinfo {title} {Attosecond physics},\
  }\href@noop {} {\bibfield  {journal} {\bibinfo  {journal} {Reviews of modern
  physics}\ }\textbf {\bibinfo {volume} {81}},\ \bibinfo {pages} {163}
  (\bibinfo {year} {2009})}\BibitemShut {NoStop}%
\bibitem [{\citenamefont {Nefedova}\ \emph {et~al.}(2017)\citenamefont
  {Nefedova}, \citenamefont {Albrecht}, \citenamefont {Kozlov{\'a}},\ and\
  \citenamefont {Nejdl}}]{nefedova2017development}%
  \BibitemOpen
  \bibfield  {author} {\bibinfo {author} {\bibfnamefont {V.}~\bibnamefont
  {Nefedova}}, \bibinfo {author} {\bibfnamefont {M.}~\bibnamefont {Albrecht}},
  \bibinfo {author} {\bibfnamefont {M.}~\bibnamefont {Kozlov{\'a}}},\ and\
  \bibinfo {author} {\bibfnamefont {J.}~\bibnamefont {Nejdl}},\ }\bibfield
  {title} {\bibinfo {title} {Development of a high-flux xuv source based on
  high-order harmonic generation},\ }\href@noop {} {\bibfield  {journal}
  {\bibinfo  {journal} {Journal of Electron Spectroscopy and Related
  Phenomena}\ }\textbf {\bibinfo {volume} {220}},\ \bibinfo {pages} {9}
  (\bibinfo {year} {2017})}\BibitemShut {NoStop}%
\bibitem [{\citenamefont {Heyl}\ \emph {et~al.}(2016)\citenamefont {Heyl},
  \citenamefont {Coudert-Alteirac}, \citenamefont {Miranda}, \citenamefont
  {Louisy}, \citenamefont {Kov{\'a}cs}, \citenamefont {Tosa}, \citenamefont
  {Balogh}, \citenamefont {Varj{\'u}}, \citenamefont {L’Huillier},
  \citenamefont {Couairon} \emph {et~al.}}]{heyl2016scale}%
  \BibitemOpen
  \bibfield  {author} {\bibinfo {author} {\bibfnamefont {C.~M.}\ \bibnamefont
  {Heyl}}, \bibinfo {author} {\bibfnamefont {H.}~\bibnamefont
  {Coudert-Alteirac}}, \bibinfo {author} {\bibfnamefont {M.}~\bibnamefont
  {Miranda}}, \bibinfo {author} {\bibfnamefont {M.}~\bibnamefont {Louisy}},
  \bibinfo {author} {\bibfnamefont {K.}~\bibnamefont {Kov{\'a}cs}}, \bibinfo
  {author} {\bibfnamefont {V.}~\bibnamefont {Tosa}}, \bibinfo {author}
  {\bibfnamefont {E.}~\bibnamefont {Balogh}}, \bibinfo {author} {\bibfnamefont
  {K.}~\bibnamefont {Varj{\'u}}}, \bibinfo {author} {\bibfnamefont
  {A.}~\bibnamefont {L’Huillier}}, \bibinfo {author} {\bibfnamefont
  {A.}~\bibnamefont {Couairon}}, \emph {et~al.},\ }\bibfield  {title} {\bibinfo
  {title} {Scale-invariant nonlinear optics in gases},\ }\href@noop {}
  {\bibfield  {journal} {\bibinfo  {journal} {Optica}\ }\textbf {\bibinfo
  {volume} {3}},\ \bibinfo {pages} {75} (\bibinfo {year} {2016})}\BibitemShut
  {NoStop}%
\bibitem [{\citenamefont {Mourou}\ \emph {et~al.}(2006)\citenamefont {Mourou},
  \citenamefont {Tajima},\ and\ \citenamefont {Bulanov}}]{mourou2006optics}%
  \BibitemOpen
  \bibfield  {author} {\bibinfo {author} {\bibfnamefont {G.~A.}\ \bibnamefont
  {Mourou}}, \bibinfo {author} {\bibfnamefont {T.}~\bibnamefont {Tajima}},\
  and\ \bibinfo {author} {\bibfnamefont {S.~V.}\ \bibnamefont {Bulanov}},\
  }\bibfield  {title} {\bibinfo {title} {Optics in the relativistic regime},\
  }\href@noop {} {\bibfield  {journal} {\bibinfo  {journal} {Reviews of modern
  physics}\ }\textbf {\bibinfo {volume} {78}},\ \bibinfo {pages} {309}
  (\bibinfo {year} {2006})}\BibitemShut {NoStop}%
\bibitem [{\citenamefont {Macchi}\ \emph {et~al.}(2005)\citenamefont {Macchi},
  \citenamefont {Cattani}, \citenamefont {Liseykina},\ and\ \citenamefont
  {Cornolti}}]{macchi2005laser}%
  \BibitemOpen
  \bibfield  {author} {\bibinfo {author} {\bibfnamefont {A.}~\bibnamefont
  {Macchi}}, \bibinfo {author} {\bibfnamefont {F.}~\bibnamefont {Cattani}},
  \bibinfo {author} {\bibfnamefont {T.~V.}\ \bibnamefont {Liseykina}},\ and\
  \bibinfo {author} {\bibfnamefont {F.}~\bibnamefont {Cornolti}},\ }\bibfield
  {title} {\bibinfo {title} {Laser acceleration of ion bunches at the front
  surface of overdense plasmas},\ }\href@noop {} {\bibfield  {journal}
  {\bibinfo  {journal} {Physical review letters}\ }\textbf {\bibinfo {volume}
  {94}},\ \bibinfo {pages} {165003} (\bibinfo {year} {2005})}\BibitemShut
  {NoStop}%
\bibitem [{\citenamefont {Esirkepov}\ \emph {et~al.}(1999)\citenamefont
  {Esirkepov}, \citenamefont {Sentoku}, \citenamefont {Mima}, \citenamefont
  {Nishihara}, \citenamefont {Califano}, \citenamefont {Pegoraro},
  \citenamefont {Naumova}, \citenamefont {Bulanov}, \citenamefont {Ueshima},
  \citenamefont {Liseikina} \emph {et~al.}}]{esirkepov1999ion}%
  \BibitemOpen
  \bibfield  {author} {\bibinfo {author} {\bibfnamefont {T.~Z.}\ \bibnamefont
  {Esirkepov}}, \bibinfo {author} {\bibfnamefont {Y.}~\bibnamefont {Sentoku}},
  \bibinfo {author} {\bibfnamefont {K.}~\bibnamefont {Mima}}, \bibinfo {author}
  {\bibfnamefont {K.}~\bibnamefont {Nishihara}}, \bibinfo {author}
  {\bibfnamefont {F.}~\bibnamefont {Califano}}, \bibinfo {author}
  {\bibfnamefont {F.}~\bibnamefont {Pegoraro}}, \bibinfo {author}
  {\bibfnamefont {N.}~\bibnamefont {Naumova}}, \bibinfo {author} {\bibfnamefont
  {S.}~\bibnamefont {Bulanov}}, \bibinfo {author} {\bibfnamefont
  {Y.}~\bibnamefont {Ueshima}}, \bibinfo {author} {\bibfnamefont
  {T.}~\bibnamefont {Liseikina}}, \emph {et~al.},\ }\bibfield  {title}
  {\bibinfo {title} {Ion acceleration by superintense laser pulses in
  plasmas},\ }\href@noop {} {\bibfield  {journal} {\bibinfo  {journal} {Journal
  of Experimental and Theoretical Physics Letters}\ }\textbf {\bibinfo {volume}
  {70}},\ \bibinfo {pages} {82} (\bibinfo {year} {1999})}\BibitemShut {NoStop}%
\bibitem [{\citenamefont {Brunel}(1987)}]{brunel1987not}%
  \BibitemOpen
  \bibfield  {author} {\bibinfo {author} {\bibfnamefont {F.}~\bibnamefont
  {Brunel}},\ }\bibfield  {title} {\bibinfo {title} {Not-so-resonant, resonant
  absorption},\ }\href@noop {} {\bibfield  {journal} {\bibinfo  {journal}
  {Physical Review Letters}\ }\textbf {\bibinfo {volume} {59}},\ \bibinfo
  {pages} {52} (\bibinfo {year} {1987})}\BibitemShut {NoStop}%
\bibitem [{\citenamefont {Gibbon}(2005)}]{gibbon2005short}%
  \BibitemOpen
  \bibfield  {author} {\bibinfo {author} {\bibfnamefont {P.}~\bibnamefont
  {Gibbon}},\ }\href@noop {} {\emph {\bibinfo {title} {Short pulse laser
  interactions with matter: an introduction}}}\ (\bibinfo  {publisher} {World
  Scientific},\ \bibinfo {year} {2005})\BibitemShut {NoStop}%
\bibitem [{\citenamefont {Nakamura}\ \emph {et~al.}(2004)\citenamefont
  {Nakamura}, \citenamefont {Kato}, \citenamefont {Nagatomo},\ and\
  \citenamefont {Mima}}]{nakamura2004surface}%
  \BibitemOpen
  \bibfield  {author} {\bibinfo {author} {\bibfnamefont {T.}~\bibnamefont
  {Nakamura}}, \bibinfo {author} {\bibfnamefont {S.}~\bibnamefont {Kato}},
  \bibinfo {author} {\bibfnamefont {H.}~\bibnamefont {Nagatomo}},\ and\
  \bibinfo {author} {\bibfnamefont {K.}~\bibnamefont {Mima}},\ }\bibfield
  {title} {\bibinfo {title} {Surface-magnetic-field and fast-electron
  current-layer formation by ultraintense laser irradiation},\ }\href@noop {}
  {\bibfield  {journal} {\bibinfo  {journal} {Physical review letters}\
  }\textbf {\bibinfo {volume} {93}},\ \bibinfo {pages} {265002} (\bibinfo
  {year} {2004})}\BibitemShut {NoStop}%
\bibitem [{\citenamefont {Buneman}(1959)}]{buneman1959dissipation}%
  \BibitemOpen
  \bibfield  {author} {\bibinfo {author} {\bibfnamefont {O.}~\bibnamefont
  {Buneman}},\ }\bibfield  {title} {\bibinfo {title} {Dissipation of currents
  in ionized media},\ }\href@noop {} {\bibfield  {journal} {\bibinfo  {journal}
  {Physical Review}\ }\textbf {\bibinfo {volume} {115}},\ \bibinfo {pages}
  {503} (\bibinfo {year} {1959})}\BibitemShut {NoStop}%
\bibitem [{\citenamefont {Bulanov}\ and\ \citenamefont
  {Sasorov}(1978)}]{bulanov1978tearing}%
  \BibitemOpen
  \bibfield  {author} {\bibinfo {author} {\bibfnamefont {S.}~\bibnamefont
  {Bulanov}}\ and\ \bibinfo {author} {\bibfnamefont {P.}~\bibnamefont
  {Sasorov}},\ }\bibfield  {title} {\bibinfo {title} {Tearing of a current
  sheet and reconnection of magnetic lines of force},\ }\href@noop {}
  {\bibfield  {journal} {\bibinfo  {journal} {Soviet Journal of Plasma
  Physics}\ }\textbf {\bibinfo {volume} {4}},\ \bibinfo {pages} {418} (\bibinfo
  {year} {1978})}\BibitemShut {NoStop}%
\bibitem [{lam()}]{lamac2022supp}%
  \BibitemOpen
  \href@noop {} {\bibinfo {title} {See \text{S}upplemental material for more
  details.}}\BibitemShut {Stop}%
\bibitem [{\citenamefont {Kahaly}\ \emph {et~al.}(2013)\citenamefont {Kahaly},
  \citenamefont {Monchoc{\'e}}, \citenamefont {Vincenti}, \citenamefont
  {Dzelzainis}, \citenamefont {Dromey}, \citenamefont {Zepf}, \citenamefont
  {Martin},\ and\ \citenamefont {Qu{\'e}r{\'e}}}]{kahaly2013direct}%
  \BibitemOpen
  \bibfield  {author} {\bibinfo {author} {\bibfnamefont {S.}~\bibnamefont
  {Kahaly}}, \bibinfo {author} {\bibfnamefont {S.}~\bibnamefont
  {Monchoc{\'e}}}, \bibinfo {author} {\bibfnamefont {H.}~\bibnamefont
  {Vincenti}}, \bibinfo {author} {\bibfnamefont {T.}~\bibnamefont
  {Dzelzainis}}, \bibinfo {author} {\bibfnamefont {B.}~\bibnamefont {Dromey}},
  \bibinfo {author} {\bibfnamefont {M.}~\bibnamefont {Zepf}}, \bibinfo {author}
  {\bibfnamefont {P.}~\bibnamefont {Martin}},\ and\ \bibinfo {author}
  {\bibfnamefont {F.}~\bibnamefont {Qu{\'e}r{\'e}}},\ }\bibfield  {title}
  {\bibinfo {title} {Direct observation of density-gradient effects in harmonic
  generation from plasma mirrors},\ }\href@noop {} {\bibfield  {journal}
  {\bibinfo  {journal} {Physical review letters}\ }\textbf {\bibinfo {volume}
  {110}},\ \bibinfo {pages} {175001} (\bibinfo {year} {2013})}\BibitemShut
  {NoStop}%
\bibitem [{\citenamefont {Dollar}\ \emph {et~al.}(2013)\citenamefont {Dollar},
  \citenamefont {Cummings}, \citenamefont {Chvykov}, \citenamefont
  {Willingale}, \citenamefont {Vargas}, \citenamefont {Yanovsky}, \citenamefont
  {Zulick}, \citenamefont {Maksimchuk}, \citenamefont {Thomas},\ and\
  \citenamefont {Krushelnick}}]{dollar2013scaling}%
  \BibitemOpen
  \bibfield  {author} {\bibinfo {author} {\bibfnamefont {F.}~\bibnamefont
  {Dollar}}, \bibinfo {author} {\bibfnamefont {P.}~\bibnamefont {Cummings}},
  \bibinfo {author} {\bibfnamefont {V.}~\bibnamefont {Chvykov}}, \bibinfo
  {author} {\bibfnamefont {L.}~\bibnamefont {Willingale}}, \bibinfo {author}
  {\bibfnamefont {M.}~\bibnamefont {Vargas}}, \bibinfo {author} {\bibfnamefont
  {V.}~\bibnamefont {Yanovsky}}, \bibinfo {author} {\bibfnamefont
  {C.}~\bibnamefont {Zulick}}, \bibinfo {author} {\bibfnamefont
  {A.}~\bibnamefont {Maksimchuk}}, \bibinfo {author} {\bibfnamefont
  {A.}~\bibnamefont {Thomas}},\ and\ \bibinfo {author} {\bibfnamefont
  {K.}~\bibnamefont {Krushelnick}},\ }\bibfield  {title} {\bibinfo {title}
  {Scaling high-order harmonic generation from laser-solid interactions to
  ultrahigh intensity},\ }\href@noop {} {\bibfield  {journal} {\bibinfo
  {journal} {Physical Review Letters}\ }\textbf {\bibinfo {volume} {110}},\
  \bibinfo {pages} {175002} (\bibinfo {year} {2013})}\BibitemShut {NoStop}%
\bibitem [{\citenamefont {Ginzburg}(1964)}]{ginzburg1966inplasma}%
  \BibitemOpen
  \bibfield  {author} {\bibinfo {author} {\bibfnamefont {V.~L.}\ \bibnamefont
  {Ginzburg}},\ }\href@noop {} {\emph {\bibinfo {title} {The Propagation of
  Electromagnetic Waves in Plasma}}}\ (\bibinfo  {publisher} {Pergamon},\
  \bibinfo {year} {1964})\BibitemShut {NoStop}%
\bibitem [{\citenamefont {Denisov}(1957)}]{deniov1957resabs}%
  \BibitemOpen
  \bibfield  {author} {\bibinfo {author} {\bibfnamefont {N.~G.}\ \bibnamefont
  {Denisov}},\ }\bibfield  {title} {\bibinfo {title} {On a singularity of the
  field of an electromagnetic wave propagated in an inhomogeneous plasma},\
  }\href@noop {} {\bibfield  {journal} {\bibinfo  {journal} {Sov. Phys. - JETP
  4}\ }\textbf {\bibinfo {volume} {4}},\ \bibinfo {pages} {544} (\bibinfo
  {year} {1957})}\BibitemShut {NoStop}%
\bibitem [{\citenamefont {Arber}\ \emph {et~al.}(2015)\citenamefont {Arber},
  \citenamefont {Bennett}, \citenamefont {Brady}, \citenamefont
  {Lawrence-Douglas}, \citenamefont {Ramsay}, \citenamefont {Sircombe},
  \citenamefont {Gillies}, \citenamefont {Evans}, \citenamefont {Schmitz},
  \citenamefont {Bell} \emph {et~al.}}]{arber2015contemporary}%
  \BibitemOpen
  \bibfield  {author} {\bibinfo {author} {\bibfnamefont {T.}~\bibnamefont
  {Arber}}, \bibinfo {author} {\bibfnamefont {K.}~\bibnamefont {Bennett}},
  \bibinfo {author} {\bibfnamefont {C.}~\bibnamefont {Brady}}, \bibinfo
  {author} {\bibfnamefont {A.}~\bibnamefont {Lawrence-Douglas}}, \bibinfo
  {author} {\bibfnamefont {M.}~\bibnamefont {Ramsay}}, \bibinfo {author}
  {\bibfnamefont {N.}~\bibnamefont {Sircombe}}, \bibinfo {author}
  {\bibfnamefont {P.}~\bibnamefont {Gillies}}, \bibinfo {author} {\bibfnamefont
  {R.}~\bibnamefont {Evans}}, \bibinfo {author} {\bibfnamefont
  {H.}~\bibnamefont {Schmitz}}, \bibinfo {author} {\bibfnamefont
  {A.}~\bibnamefont {Bell}}, \emph {et~al.},\ }\bibfield  {title} {\bibinfo
  {title} {Contemporary particle-in-cell approach to laser-plasma modelling},\
  }\href@noop {} {\bibfield  {journal} {\bibinfo  {journal} {Plasma Physics and
  Controlled Fusion}\ }\textbf {\bibinfo {volume} {57}},\ \bibinfo {pages}
  {113001} (\bibinfo {year} {2015})}\BibitemShut {NoStop}%
\bibitem [{\citenamefont {Jackson}(1999)}]{jackson1999classical}%
  \BibitemOpen
  \bibfield  {author} {\bibinfo {author} {\bibfnamefont {J.~D.}\ \bibnamefont
  {Jackson}},\ }\href@noop {} {\emph {\bibinfo {title} {Classical
  electrodynamics}}}\ (\bibinfo  {publisher} {John Wiley \& Sons},\ \bibinfo
  {year} {1999})\BibitemShut {NoStop}%
\bibitem [{\citenamefont {Hirschmugl}\ \emph {et~al.}(1991)\citenamefont
  {Hirschmugl}, \citenamefont {Sagurton},\ and\ \citenamefont
  {Williams}}]{hirschmugl1991multiparticle}%
  \BibitemOpen
  \bibfield  {author} {\bibinfo {author} {\bibfnamefont {C.~J.}\ \bibnamefont
  {Hirschmugl}}, \bibinfo {author} {\bibfnamefont {M.}~\bibnamefont
  {Sagurton}},\ and\ \bibinfo {author} {\bibfnamefont {G.~P.}\ \bibnamefont
  {Williams}},\ }\bibfield  {title} {\bibinfo {title} {Multiparticle coherence
  calculations for synchrotron-radiation emission},\ }\href@noop {} {\bibfield
  {journal} {\bibinfo  {journal} {Physical Review A}\ }\textbf {\bibinfo
  {volume} {44}},\ \bibinfo {pages} {1316} (\bibinfo {year}
  {1991})}\BibitemShut {NoStop}%
\end{thebibliography}%

\end{document}